\documentclass[usenatbib,a4paper,11pt]{article}
\pdfoutput=1 
\usepackage{jcappub}
\usepackage{soul}
\usepackage[T1]{fontenc}
\usepackage{subcaption}
\usepackage[dvipsnames]{xcolor}
\usepackage{soul}
\usepackage{hyperref}
\usepackage{CJK}
\usepackage{multirow}
\usepackage{booktabs}
\usepackage{color}

\newcommand{\Mpch}{\mathinner{{\rm Mpc}/h}}
\newcommand{\hMpc}{\mathinner{h/{\rm Mpc}}}
\newcommand{\Gpch}{\mathinner{{\rm Gpc}/h}}
\newcommand{\Msunh}{\mathinner{{\rm M_\odot}/h}}

\newcommand{\Om}{\mathinner{\Omega_{\rm m}}}

\newcommand{\ns}{\mathinner{n_{\rm s}}}

\title{\boldmath The Universe is worth $64^3$ pixels: Convolution Neural Network and Vision Transformers for Cosmology }

\author[a,b]{Se Yeon Hwang,}
\author[a,b]{Cristiano G. Sabiu,\footnote{Corresponding Author}}
\author[a,b]{Inkyu Park,}
\author[c,d]{Sungwook E. Hong}

\affiliation[a]{Department of Physics, University of Seoul, 163 Seoulsiripdae-ro, Dongdaemun-gu, Seoul 02504, Republic of Korea}
\affiliation[b]{Natural Science Research Institute, University of Seoul, 163 Seoulsiripdae-ro, Dongdaemun-gu, Seoul 02504, Republic of Korea}
\affiliation[c]{Korea Astronomy and Space Science Institute, 776 Daedeok-daero, Yuseong-gu, Daejeon 34055, Republic of Korea}
\affiliation[d]{Astronomy Campus, University of Science and Technology, 776 Daedeok-daero, Yuseong-gu, Daejeon 34055, Republic of Korea}

\emailAdd{worldhsy@gmail.com}
\emailAdd{csabiu@uos.ac.kr}
\emailAdd{icpark@uos.ac.kr}
\emailAdd{swhong@kasi.re.kr}

\abstract{
We present a novel approach for estimating cosmological parameters, $\Omega_m$, $\sigma_8$, $w_0$, and one derived parameter, $S_8$, from 3D lightcone data of dark matter halos in redshift space covering a sky area of $40^\circ \times 40^\circ$ and redshift range of $0.3 < z < 0.8$, binned to $64^3$ voxels. Using two deep learning algorithms—Convolutional Neural Network (CNN) and Vision Transformer (ViT)—we compare their performance with the standard two-point correlation (2pcf) function. Our results indicate that CNN yields the best performance, while ViT also demonstrates significant potential in predicting cosmological parameters. By combining the outcomes of Vision Transformer, Convolution Neural Network, and 2pcf, we achieved a substantial reduction in error compared to the 2pcf alone. To better understand the inner workings of the machine learning algorithms, we employed the Grad-CAM method to investigate the sources of essential information in heatmaps of the CNN and ViT. 
Our findings suggest that the algorithms focus on different parts of the density field and redshift depending on which parameter they are predicting.
This proof-of-concept work paves the way for incorporating deep learning methods to estimate cosmological parameters from large-scale structures, potentially leading to tighter constraints and improved understanding of the Universe.}

\begin{document}
\maketitle
\flushbottom

\section{Introduction}
\label{sec:intro}
The large-scale structure (LSS) of the universe offers invaluable insights into the underlying cosmological model. The distribution of galaxies and dark matter in the LSS is governed by the initial conditions and cosmological parameters that dictate the evolution of the universe \citep{1980lssu.book.....P,1985ApJ...292..371D}. Thus, comparison between observational and theoretical models or numerical simulations provides constraints on the components and physics that govern the Universe on the largest scales.
Visually, the LSS of galaxies is a complex web-like distribution exhibiting over-dense ``clusters'', near-empty ``void'' regions and intermediate density features known as ``walls'' and ``filaments'' \cite{1996Natur.380..603B}. These structures are also not static but change and evolve over cosmic time, or redshift, due to gravity and cosmic expansion. However, to compare observations with theoretical models, we typically condense this rich 3-dimensional galaxy distribution into a 2nd-order clustering statistic such as the power spectrum (PS) or correlation function (CF), although there are some efforts to move into higher order statistical descriptions of the LSS \cite{2016A&A...592A..38S,2019ApJS..242...29S,2021arXiv210801670P}.

Measurements of the galaxy PS or CF have provided exquisite constraints on the expansion history of the Universe via the Baryon Acoustic Oscillations (BAO) \cite{Eisenstein_2005, 2005MNRAS.362..505C, 2011MNRAS.416.3017B} and the Alcock-Paczynski effect \cite{2019ApJ...887..125L,2019ApJ...881..146P,2019ApJ...878..137Z,2018ApJ...856...88L,2017ApJ...844...91L,2016ApJ...832..103L,2015MNRAS.450..807L}, and on the the growth of structure via Redshift Space Distortions (RSD) \cite{1987MNRAS.227....1K, 2017MNRAS.470.2617A,2021MNRAS.500..736B,2021MNRAS.501.5616D,2020MNRAS.499..210N}. In many of these works, the observed statistic are compared with analytic models, templates or more numerical perturbation theory predictions. However, in an effort to reliably predict the clustering on small scales, cosmological $N$-body simulations have been employed \cite{2020ApJS..250....2V, 2022MNRAS.515..871Y,2023MNRAS.520.6283Y,2023JCAP...02..050K}. Also, using a suite of simulations, one can build emulators predicting clustering statistics via efficient sampling and interpolation, for a range of cosmological models and parameters \cite{Hahn_2023,2022PhRvD.105h3517K}.  
Although numerically expensive, this approach has many advantages, including easy applications of observational systematics. 

In recent years, machine learning techniques have demonstrated tremendous potential for extracting information from large and complex datasets in various scientific disciplines \cite{2015Natur.521..436L,GoodBengCour16}. In the context of cosmology, recently, Convolution Neural Networks (CNNs) have shown promise in predicting cosmological parameters from the LSS density field \cite{Ravanbakhsh, Mathuriya, pan, Lazanu, Ntampaka_2020} and from 21cm temperature maps at higher redshift \cite{2022JCAP...01..020S}. One advantage of this technique is that it allows us to directly use the distribution of matter without the need of somewhat processed statistics. In ref.~\cite{Ravanbakhsh}, they first use a 3D CNN with $N$-body simulations to predict the matter density parameter $\Om$ and the root-mean-square of the amplitude of matter perturbation at $8\Mpch$-scale $\sigma_8$. 
In ref.~\cite{Mathuriya}, they used a similar model structure as ref.~\cite{Ravanbakhsh}
to predict $\Om$, $\sigma_8$ and the scalar spectral index of the primordial PS $\ns$. In ref.~\cite{pan}, they leveraged these pioneering works in a more lightweight approach with smaller input size to predict $\Om$ and $\sigma_8$. In ref.~\cite{Lazanu}, while they predicted $\Om$ and $\sigma_8$ using a 3D CNN, they were also able to predict $\ns$ by combining with PS. In ref.~\cite{Ntampaka_2020}, by using  mock galaxy simulations with various Halo Occupation Distributions (HOD), they predicted $\Om$ and $\sigma_8$. Alternatively, cosmological parameters can also be predicted using the projection of 3D data, as demonstrated by ref.~\cite{deeplss}. The examples presented here are summarized in table~\ref{table1}.

\begin{table}[tbp]
\centering
\begin{tabular}{cccc
}
\toprule
Literature  & Data Type  & Data Size                          
& Prediction  \\ \toprule
\cite{Ravanbakhsh}  
& Halo, Snapshot ($z = 0$)  & $64^3$, $(128\Mpch)^3$  &  
$\Om, \sigma_8$     \\ \midrule
\cite{Mathuriya}   
& Halo, Snapshot ($z = 0$)  & $128^3$, $(256\Mpch)^3$  
& $\Om, \sigma_8, \ns$ \\ \midrule
\cite{pan}   
& DM Density, Snapshot ($z = 0$)  & $32^3$, $(64\Mpch)^3$   
& $\Om, \sigma_8$     \\ \midrule
\cite{Lazanu}  
& DM Density, Snapshot ($z = 0$)  & $64^3$, $(1\Gpch)^3$                            
& $\Om, \sigma_8$        \\ \midrule
\multirow{2}{*}{\cite{Ntampaka_2020}} & \multirow{2}{*}{HOD, Snapshot ($z = 0.3$)} & $275 \times 275 \times 55$, & \multirow{2}{*}{$\Om, \sigma_8$} \\
& & $550 \times 550 \times 220~(\Mpch)^3$ & \\ \midrule
\multirow{2}{*}{Our Study} & Halo, Lightcone & $64^3$, $(2\Gpch)^3$ & \multirow{2}{*}{$\Om, \sigma_8, w_0$} \\ 
& (0.3 < z < 0.8) & $40^\circ \times 40^\circ$ &  \\ \bottomrule
\end{tabular}
\caption{\label{table1} Summary of the cosmological parameter predictions using 3D CNN with simulations described in the texts.}
\end{table}

While CNNs have been widely used in astronomy, Vision Transformer (ViT) emerged from computer vision tasks \cite{vit}, and the attention mechanism \cite{attention} underlying them was widely used in Natural Language processing. While CNN reduces the data size by half using a pooling layer after the convolution layer, 
ViT maintains the original data size. ViT divides the 2D or 3D image into small patches and flatten to 1D array to calculates scores indicating how much attention should be given between each patches. This approach allows us to identify the important patches and extract the desired features from the data. Furthermore, they are suitable for finding global relationships between each patch. Several studies have utilized ViT in the field of astronomy. In ref.~\cite{vit_gal_morphology}, they used ViT to solve galaxy morphological classification with the Galaxy Zoo dataset.  
They found that, although the total accuracy means of CNN were higher than ViT, it performed better in classifying smaller and fainter galaxies than CNN. In ref.~\cite{vit_unsupervised_galmorp}, galaxy classifications were also studied using  
Galaxy Zoo 2, Sloan Digital Sky Survey Data Release 17 (SDSS-DR17), and Galaxy Zoo Dark Energy Camera Legacy Survey (DECaLS).  
They found that using the combination of CNN and ViT had the best performance, followed by the CNN and the ViT model. In ref.~\cite{vit_strong_lensing}, they predicted strong gravitational lensing parameters with ViT using 31,200 simulated strongly lensed quasar images, 
where the ViT model outperformed the CNN model  except for one parameter. Lastly, ref.~\cite{vit_wgan} tested the morphological classification of AI-augmented images of radio galaxies, where a Fully-Connected Neural Network (FCN) model outperforms both CNN and ViT. 
In some cases, ViT has competitive performances with CNN especially when they are pre-trained with large data set. On the other hand, in this paper, we apply ViT for the first time in predicting cosmological parameters using 3d simulations without pre-training and compare them with CNN results.

The remainder of this paper is organized as follows:  Section~\ref{sec:data} describes the data which is used to train and test our model and also the data which we used for making two point correlation functions. In Section~\ref{sec3}, we introduce the machine learning models that we test in this work, both for Convolution Neural Network and Vision Transformer. In Section~\ref{sec:analysis}, we analysis the result from CNN, ViT and from the two-point correlation function. We discuss how we can interpret and understand the procedure of deep learning and which information was valuable when deep learning predicts the cosmological parameters. Lastly, we make conclusions in Section~\ref{sec:conc}.

\section{Data}
\label{sec:data}

\subsection{Cosmological Parameter Sets}\label{sec:data_cosmo}

By assuming flat $w$CDM cosmologies, we vary five cosmological parameters within specified ranges: $\Om \in [0.25, 0.4]$, $\sigma_8 \in [0.4, 1.1]$, the equation-of-state of dark energy $w_0 \in [-1.5, -0.5]$, $\ns \in [0.9, 1.1]$, and $h \in [0.6, 0.8]$. Then, we calculate the derived matter perturbation amplitude $S_8 \equiv \sigma_8 (\Om/0.3)^{0.5}$. We adopt Latin Hypercube Sampling to select 994 distinct parameter sets for our simulation suite.

Our datasets can be categorized into two types. First, training (80\% of total datasets) and test (20\% of total datasets) datasets comprise different cosmological parameters and initial random seeds. Then, a systematic test set assesses the performance of the trained model on a single, fixed cosmological parameter set, where only the initial seeds are varied. For this systematic test set, we fix the cosmological parameters as $(\Om, \sigma_8, w_0, \ns, h) = (0.3133, 0.8079, -1, 0.9649, 0.6736)$, which is in a concordance with the flat $\Lambda$CDM cosmology from Planck 2018 \cite{planck2018}, and generate 100 single cosmology datasets. 

\subsection{Mock Lightcone Halo Catalogs}\label{sec:data_halo}

In predicting cosmological parameters, it is crucial to consider the application of observational data. The ultimate goal is to predict more accurate parameters from observational data. As summarized in Table~\ref{table1}, our study's distinguishing feature is the use of lightcone data with a broad redshift range. 
This enabled us to try predicting the dark energy parameter $w_0$ for the first time using the 3D CNN method.

For each cosmological parameter set from the above subsection, we employ the 
PINpointing Orbit Crossing Collapsed HIerarchical Objects ({\tt PINOCCHIO}) algorithm \cite{Monaco_2002} to generate a mock lightcone dark matter (DM) halo catalogs. {\tt PINOCCHIO} uses Lagrangian Perturbation Theory (LPT) to produce DM halo catalogs containing halo mass, position, and velocity information. 
Each simulation was performed with a comoving box size of $2\Gpch$ and $1024^3$ DM particles. Then we utilize the automatic past lightcone data generation of the {\tt PINOCCHIO} code, which assigns the right ascension (RA), declination (DEC), and observed redshift ($z$) for each DM halo.
We set the output range of RA, DEC, and redshift to have a complete past lightcone halo catalog within $-20^\circ < {\rm RA, DEC} < +20^\circ$ and $0.3 < z < 0.8$. 

The halo mass function strongly depends on cosmological parameters \citep{1974ApJ...187..425P, 1999MNRAS.308..119S, 2001MNRAS.321..372J, 2008ApJ...688..709T}. Consequently, the number of halos in each simulation inherently provides some amount of cosmological information, which we aim to utilize. However, the minimum halo mass for each simulation, related to the mass of individual DM particles, depends crucially on $\Om$. Therefore, simulations with different values of $\Om$ probe different regimes of the halo mass function, resulting in significantly different values for the mean halo number density. 
We want to retain the former effect while eliminating the latter. To achieve this, we use the largest minimum halo mass across our simulation suite $M_{\rm min} = 8.27\times10^{12} \Msunh$ to impose a mass cut on all simulations. This would allow us to compare more directly to observations where a faint magnitude or luminosity cut is naturally imposed, although we leave such comparisons for future work.
Our range of mean halo number density for different cosmology sets vary from $3.29 \times 10^{-5} (\hMpc)^3$ to $6.88 \times 10^{-4} (\hMpc)^3$. On the other hand, the mean halo number density from 100 single cosmology datasets is $(3.9 \pm 0.017) \times 10^{-4} (\hMpc)^3$, which is close with galaxy number density from SDSS-III BOSS CMASS ($\sim 4\times 10^{-4} (\hMpc)^3$). This CMASS sample targets Luminous Red Galaxies (LRGs) at 0.4 < z < 0.7, which is similar to our redshift range. Additionally most LRGs are expected to be around the center of relatively massive clusters. Since our DM halos have a similar number density to CMASS galaxies, we consider that this data would be appropriate to approximately mimic CMASS-like galaxies, with the caveat that we are imposing a fixed halo occupation scheme.

The mock simulations also do not consider observational effects such as fiber collision or complicated angular masking. However, this simplification affects both our 2pcf and machine learning analyses. So while these mock simulations cannot immediately be compared to observational samples, we expect that they are complex enough to compare the cosmological information content between ML and traditional statistics, which is the primary goal of our study.

\subsection{Input Data for Machine Learning \& Two-point Correlation Function}\label{sec:data_input}

For making the input data for the 3D CNN and ViT deep learning methods, we divide the RA and DEC into two segments each: $(-20^\circ, 0^\circ)$ and $(0^\circ, +20^\circ)$, which leads to a fourfold increase in the number of samples. Then we bin the DM halo positions of each lightcone to $64^3$ voxels, each axis consists of 64 bins, with one bin representing 0.3125 degrees in RA and DEC, while containing different comoving distances. Before feeding the data into deep learning algorithms, we normalize the values by dividing the total maximum number of DM halos in a single voxel from all 994 datasets.

As a comparison to the deep learning methods, we compute the isotropic two-point correlation function $\xi(r)$ using the public code \texttt{KSTAT} \cite{2018ascl.soft04026S} with the Landy \& Szalay estimatior \cite{LS1993}. 
To properly include the expansion history, we divide each lightcone halo catalog into three redshift slices: $0.3 < z < 0.5$, $0.5 < z < 0.65$, and $0.65 < z < 0.8$, which roughly contain a similar number of DM halos. We found that three is the optimal number of redshift bins to allow constraints on $w_0$ without degrading the constraining power for other parameters. 
Then, we calculate the 2pcf from each redshift shell and concatenate them.  
To avoid the Finger-of-God effect, we calculate the 2pcf from $5 - 150\Mpch$ with 29 bins, with each bin step size $5\Mpch$. We further remove the first two radial bins because they are smaller than the cell size of our density field in the deep learning approach and they may otherwise give an unfair advantage to the 2pcf results.

\section{Machine Learning Models}
\label{sec3}

Convolutional Neural Networks (CNN) are a prevalent methodology for machine learning tasks in astronomy, especially for computer vision. However, after the introduction of the Vision Transformer (ViT) mechanism, we have found potential in using ViT for predicting cosmological parameters. The main difference between CNN and ViT is related to their inductive bias, which is the ability to predict unseen images during training. ViT has a lower inductive bias than CNN, and that is why training ViT with large datasets is crucial for achieving good performance. In \cite{vit}, ViT outperformed CNN when trained with a large dataset containing 300 million images.  
However, producing large datasets in astronomy through simulations has limitations in generating simulations, saving it and also running deep learning algorithms. If one have not enough astronomical data, they can use pre-trained ViT models with other datasets \cite{vit_strong_lensing, vit_wgan}.  
However, we do not use a pre-trained model and use the same number of data for both CNN and ViT. 
During training we apply flipping augmentation in the angular plane, while preserving the redshift axis to retain the correct large scale structure evolution.
The code is implemented using {\tt Tensorflow} \cite{tensorflow2015-whitepaper} and one NVIDIA RTX 3090 graphic card unit (GPU) with 24~GB VRAM. For the loss function, we use the Mean Squared Error (MSE),

\begin{equation}
\label{eq1}
\mathcal{L} = \frac{1}{N_{\rm batch}}\sum_{i=1}^{N_{\rm batch}} \sum_{j=1}^{4} \left( x_{j,i}^{\rm pred} - x_{j,i}^{\rm truth} \right)^2 ~,
\end{equation}
where $(x_1, x_2, x_3, x_4) = (\Om, \sigma_8, w_0, S_8)$, and $N_{\rm batch}$ is the number of data in each minibatch.
We tried different weights which multiplied each parameters errors, but we couldn't find any advantage of this and decided not to use weights in our loss function.
Also, to check the convergence of each machine learning model, we perform 10 independent runs to get the results.

\subsection{Convolutional Neural Networks}

Main part of CNN consists of the convolution layer and the pooling layer. During the convolution layer, a small kernal moves around the entire image and performs convolution operations with each value. After the convolution layer, we apply the pooling layer, which reduces the data size by half. There are two types of pooling layer: average pooling and maximum pooling. The average pooling calculates the average value of pixels, while the maximum pooling selects the maximum value of pixels. 
We use average pooling in this paper, while maximum pooling also produced similar results in our tests.

\begin{figure}[tbp]
\centering \includegraphics[width=\textwidth]{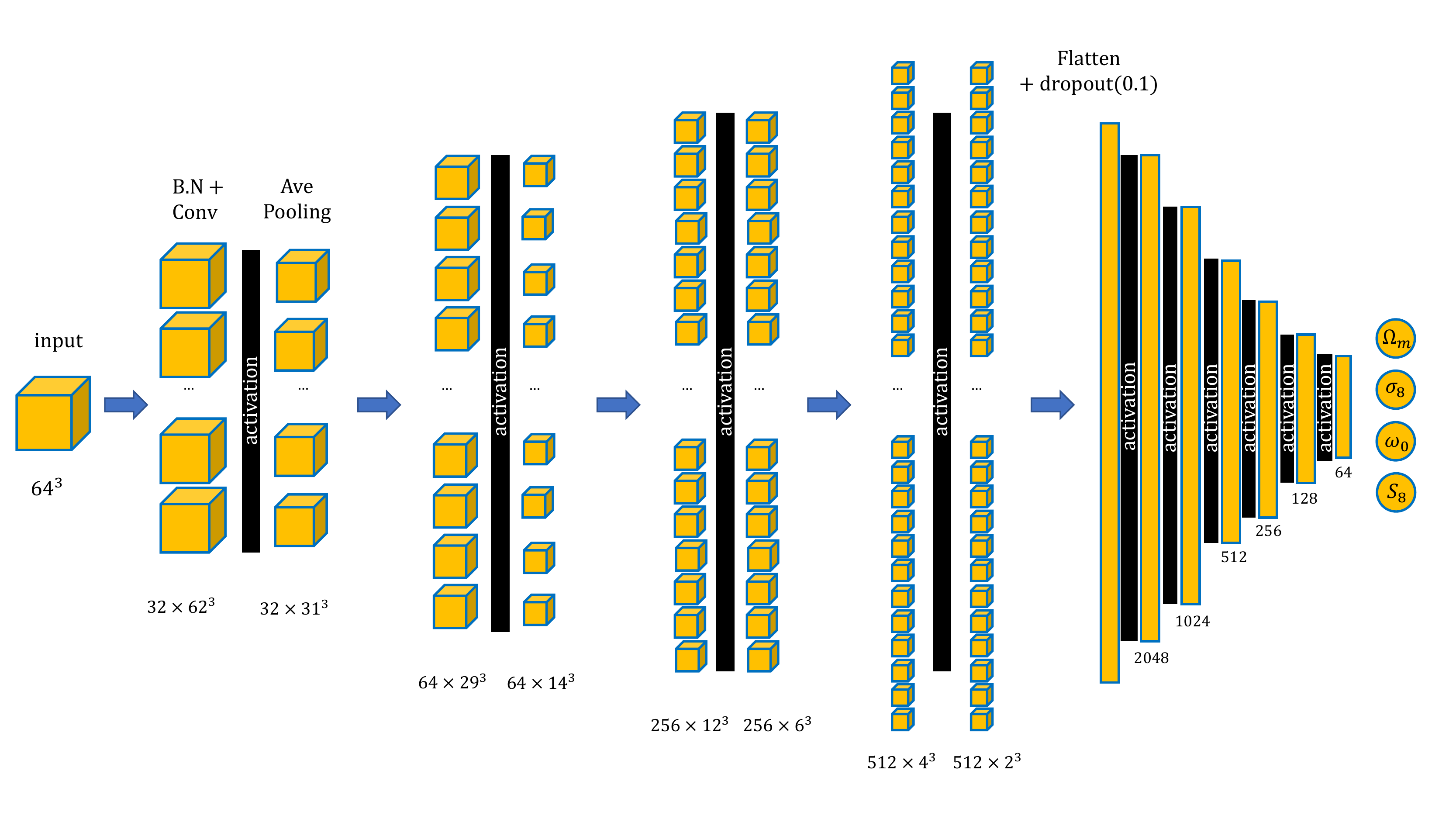}
\caption{\label{fig1} CNN structure of this work. 
4 Convolution layers with filters 32, 64, 256 and 512 have applied for feature extraction, followed by 7 fully connected layers with 2,048, 1,024, 512, 256, 128, 64 and 4 neurons for parameter estimation.}
\end{figure}

Figure~\ref{fig1} shows our CNN model, which consists of four convolution layers with  32, 64, 256, and 512 filters. For kernel size, we used $3^3$ with strides for (1,1,1) and no padding. We put a batch normalization \cite{2015arXiv150203167I} layer before the convolution layer to prevent the vanishing gradient problem. 
After the convolution layer, we sequentially put an activation layer and an average pooling layer. After finishing the convolution layer, we flatten the output data into 1D data and apply 10\% dropout layer \cite{srivastava2014dropout} to prevent the overfitting problem. Then, we put seven FCN layers with 2,048, 1,024, 512, 256, 128, 64, and 4 neurons, each of which followed by batch normalization and activation. 
Every activation function used in the convolution layer and dense layer is a Rectified Linear Unit (ReLU, \cite{glorot2011deep}) ${\rm ReLU}(x) = \max (x, 0)$, except for the last two activation functions, which are linearly activated. {This model has a total of 15,244,200 trainable parameters.

We use the Adam optimizer \cite{kingma2014adam}, a batch size of 64, and a learning rate of 1e-05. 
We train our CNN model until epoch to 700, which takes about 106 minutes with a single NVIDIA RTX 3090 GPU. We use 10\% of the training data for validation and adopt the model at the epoch where the validation loss is minimum.

\subsection{Vision Transformers}
\label{sec:vitmodel}

\begin{figure}[tbp]
\centering 
\includegraphics[width=\textwidth]{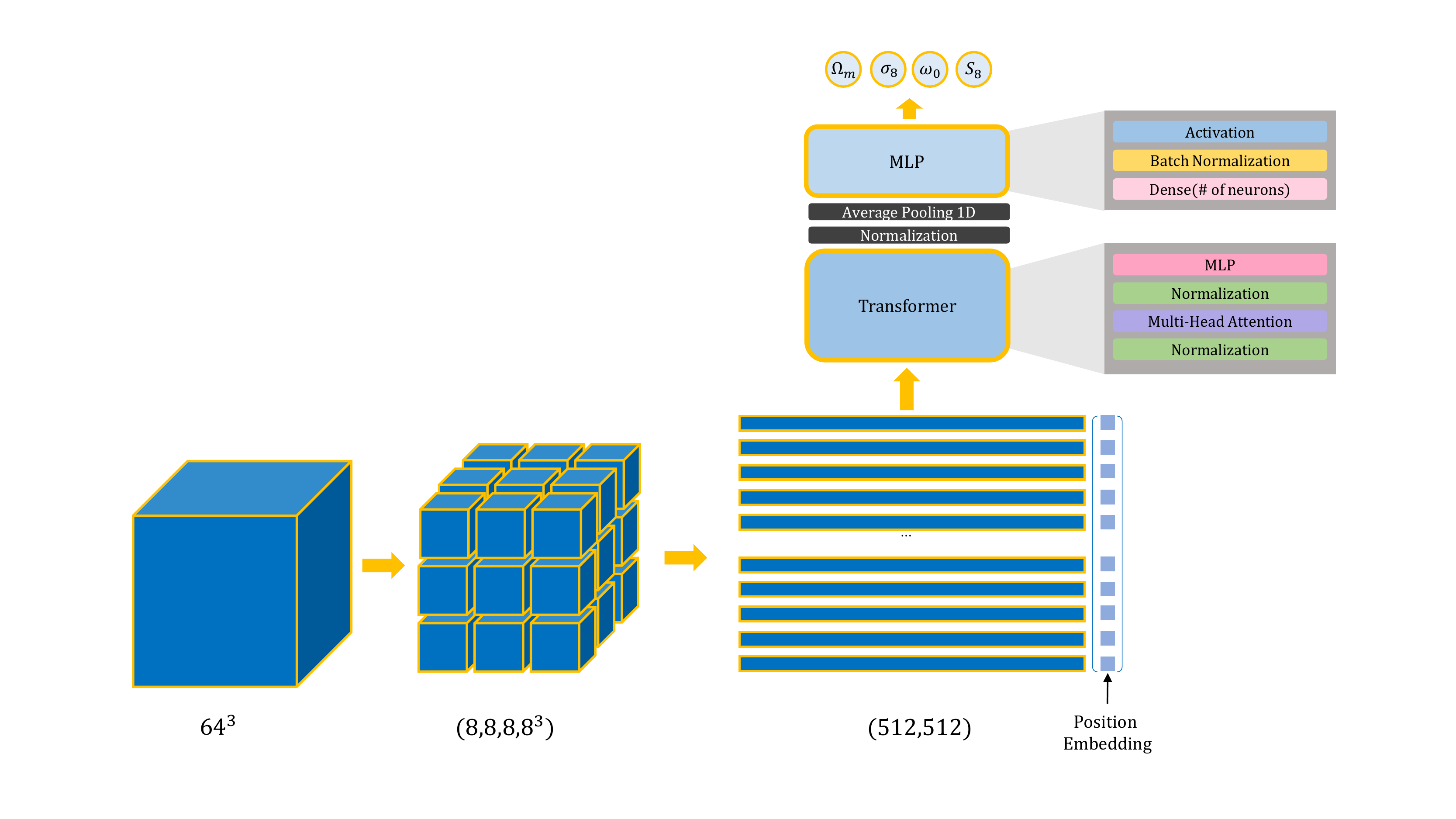}
\caption{\label{fig2} ViT structure of this work. We divide a single $64^3$ data into (8,8,8) patches, which indicate patches position $(x,y,z)$, where a single patch is a $8^3$ cube, and flatten to a 1D array and added to position embedding vectors.
It then undergoes 4 heads in Multi-Head Attention, with a layer depth of 4. After attention layers, we apply the Multi-Layer Perceptron (MLP), which is one Fully Connected layer. }
\end{figure}

Vision Transformers (ViT) are a novel approach to computer vision tasks, which adapt the Transformer architecture, originally developed for Natural Language Processing (NLP), to handle image-like data \cite{vit}. In the context of this work, we can employ ViT to analyze the spatial relationships between different parts of the input data, which consist of flattened 3D patches extracted from the lightcone simulation with an input size of $64^3$.

Our specific implementation of ViT, as illustrated in Figure~\ref{fig2}, divides the input data into $8^3$ non-overlapping patches, resulting in a total of 512 patches. These patches are then passed through an embedding layer with a projection dimension of 512. Note that we additionally incorporate positional encoding to convey the relative positions of the tokens in the input sequence. This is crucial for ViT because the spatial relationships between the patches are essential for understanding the structure of the input data.

The main building block of the Transformer architecture is the self-attention mechanism. This mechanism allows the model to learn the relationships between different tokens in the input sequence. In our case, the tokens are the flattened 3D patches. The self-attention mechanism operates on three components: Queries ($Q$), Keys ($K$), and Values ($V$). 
To create the $Q$, $K$, and $V$, the flattened patches are first projected through learnable linear transformations. These transformations produce three weight matrices, $W_K$, $W_Q$, and $W_V$. Then by multiplying the projected patches with weight matrices, we can finally obtain $Q$, $K$ and $V$. 
The $Q$ and $K$ are used to calculate the attention scores between patches, while the $V$ is used to compute the output patch embeddings. The attention scores are computed as the dot product between the $Q$ and the $K$, scaled by the square root of the dimensionality of the $K$ vectors and multiplied by $V$.

Next, we apply the Transformer Encoder, a series of Multi-Head Attention (MHA) layers, each of which utilizes multiple attention scores from a given (Queries, Keys, Values) set.
First, a single attention score is defined as \cite{attention}

\begin{equation}
\label{eq3.2}
    {\rm Attention} = {\rm softmax} \left(\frac{Q K^{\rm T}}{\sqrt{d_K}}\right) V ~,
\end{equation}
where the superscript ``T'' denotes the matrix transpose, and $d_K$ is the dimension of $K$-matrix.
Then each MHA layer is defined as 
\begin{equation}
    {\rm MultiHead}(Q,K,V) = {\rm Concat}({\rm head}_1, \ldots, {\rm head}_h)W^{O} ~,
\end{equation}
where ${\rm head}_i = {\rm Attention}(QW_i^Q, KW_i^K, VW_i^V)$ is the $i$-th attention score, and $\{W_i^Q, W_i^K, W_i^V\}$ and $W^O$ are weight matrices that will be learned throughout training.
This process of computing attention scores and generating output embeddings is performed multiple times, in parallel, by different ViT heads in order to capture different aspects of the image. The resulting output embeddings from all heads are then concatenated and passed through a feedforward neural network to produce the final output.

After the Transformer Encoder processes the input patches, we normalize the outputs to maintain the mean within each batch close to 0 and the standard deviation close to 1. Subsequently, we apply global average pooling to obtain 1D layers from the Transformer outputs. After adding a 10\% dropout layer, we apply a series of fully connected (FCN) layers with 2,048, 1,024, 512, 256, 128, 64, and 4 neurons. To ensure a fair comparison with the CNN model, we use the same structure after the Transformer layer. Note that we did not include any dropout layer inside the FCN layers, as this does not improve the results. This model has a total of 25,425,348 trainable parameters.


While training we use the Adam optimizer, a learning rate of 5e-06, and a batch size of 32 due to memory limitations. This took about 226 minutes on the same GPU as before and again we use the model with the minimum validation loss. In creating this model, we modified the publically available \texttt{Keras} example code\footnote{\url{https://keras.io/examples/vision/image\_classification\_with\_vision\_transformer}}.

\subsection{Two-Point Correlation Function Neural Network}
\label{sec3.3}

We train Fully Connected Layers (FCN) 
to utilize the 2pcf measurements done in Section~\ref{sec:data_input} to estimate cosmological parameters, which we call ``2pcf+FCN'' hereafter. 
We first flatten our input data shape of (27,3), where 27 indicates 2pcf radial bins ($15-150\Mpch$) and 3 is the number of redshift shells ($0.3 < z < 0.5$, $0.5 < z < 0.65$, and $0.65 < z < 0.8$), giving us an input vector of 81 elements. We pass this into a 512-neuron hidden layer and then perform batch normalization and activated with a ReLU function. Next, we pass them through a 32-neuron hidden layer and linear activation. Lastly we pass the output from the previous layer to a 4-neuron output layer. We also tried more complex network structures and found that the current version performs the best.

For this running, we choose the learning rate of Adam optimizer to 5e-05 and the number of epochs to 500 to obtain stable results.
Compared to our previous CNN and ViT models, the 2pcf+FCN model takes a negligible amount of time for training, while the calculation of 2pcf data takes more than an hour.

\section{Analysis}
\label{sec:analysis}

In this section, we discuss the results obtained from two deep learning algorithms, CNN and ViT, and compare their performance with the 2pcf + FCN. Section~\ref{sec4.1} presents a comparison of the outcomes from the three methods and discuss the test results for different cosmological sets. We also examine the results from the systematic test dataset. 
In Section~\ref{sec4.2}, we explore the correlation between the three methods using Pearson correlation and present the combined results from all three approaches. Finally, in Section~\ref{sec4.3}, we try to understand the deep learning procedures by analyzing the critical information extracted from the data.

\subsection{Model Comparison}
\label{sec4.1}

We begin by comparing the results obtained from the three methods, CNN, ViT, and 2pcf+FCN, for each cosmological parameter ($\Om, \sigma_8, w_0, S_8$). In the context of different cosmological sets, we analyze the performance of these methods in terms of accuracy and consistency. Furthermore, we investigate the impact of varying initial random seeds on the outcomes by performing a single cosmological test with 100 variations.
For our result, we use the output from 10 runs to reduce the inherent randomness in the initialization process of neural networks and to avoid relying on results from only one run which may yield conclusions that are sensitive to the specific initialization. By conducting 10 runs, we generate a spread of outcomes that, when analyzed collectively, mitigate the uncertainties arising from these random initializations. This provides a more robust and credible understanding of the machine learning model's performance. In machine learning this is commonly referred to as model averaging.
\begin{figure}[tbp]
\centering
\includegraphics[width=\textwidth]{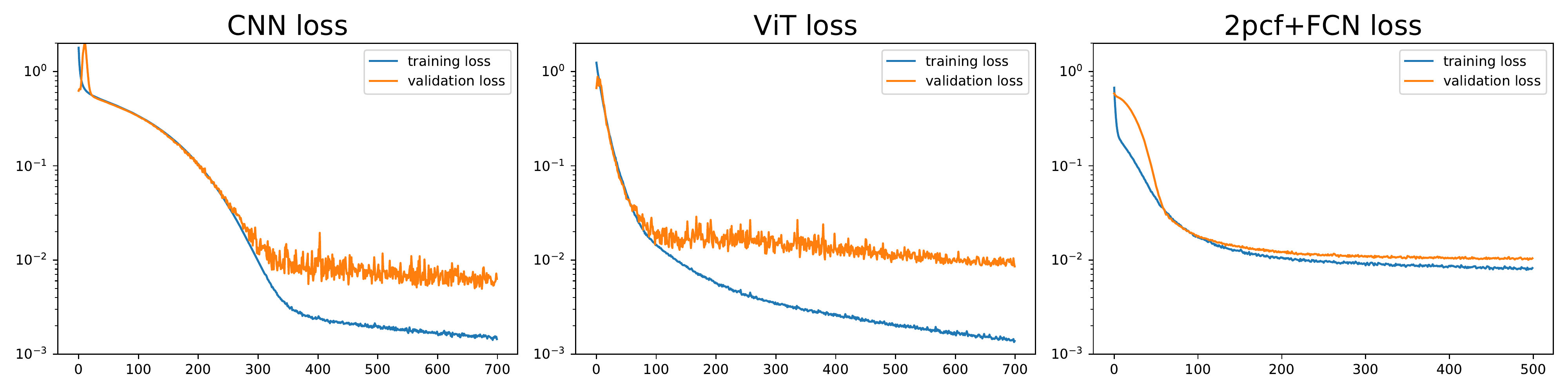}
\caption{This is 10 averaged loss curve from 10 running sets of CNN (left), ViT (middle), and 2pcf+FCN (right) methods. The blue line is the training loss and the orange line is the validation loss.}
\label{fig3}
\end{figure}
 
Figure~\ref{fig3} presents the loss curves averaged value, at the same epoch, from 10 independent runs of our three methods. Among the training data, which accounts for 80\% of the total dataset, we allocated 90\% data to the training loss and 10\% data to the validation loss. We checked the loss curves for each individual run and there were no outliers.  By doing the averaging, we aimed to make comparisons on the same level as the other results presented here. Since we chose the model weights at the lowest validation loss for our final model, it is worth to compare their minimum validation loss. When we identify the lowest validation loss from each run and average the total of 10 values, the order is CNN (0.00404 $\pm$ 0.00032), ViT (0.00757 $\pm$ 0.00048), and 2pcf+FCN (0.00973 $\pm$ 0.00185), where standard deviation comes from 10 runs. The minimum validation loss aligns well with our final result.

\begin{figure}[tbp]
\centering 
\includegraphics[height=0.85\textheight]{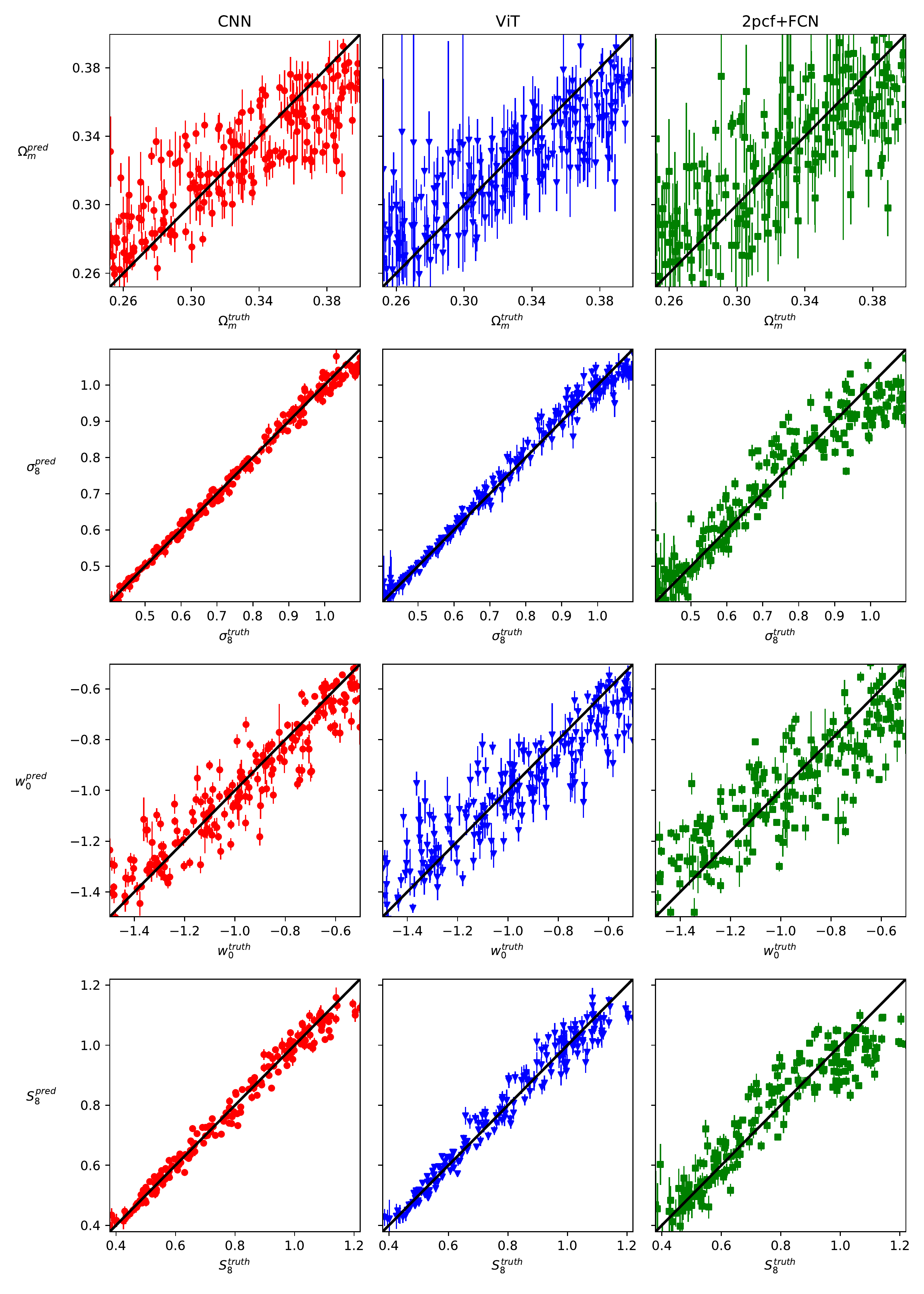}
\caption{\label{fig4} Comparison between truth values of $\Om, \sigma_8, w_0$, and $S_8$ (top to bottom) from test samples and their predictions from CNN (left), ViT (middle), and 2pcf+FCN (right). Dots and error bars correspond to average and standard deviation from 10 different runs, respectively. Black diagonal line shows where the prediction is same to the truth. 
}
\end{figure}

Figure~\ref{fig4} presents the one-to-one comparison between truth and prediction values of $(\Om, \sigma_8, w_0, S_8)$ from the test data.  
Because we have four subcubes from one cosmological parameter set, we average the prediction from each subcube and use that for our final points. Also, to validate our deep learning model, we run the model 10 times under the same structure and plot the average value and standard deviation for error. All three models can predict all four cosmological parameters, while there are some differences between models and parameters. The predictions of 
$\Om$ and $w_0$ seem to be more scattered and have larger error bars than the other parameters in all three models, mainly because their allowed ranges are narrower than others.

In Figure \ref{fig5} we show the marginalised 2D contours from the single cosmology dataset with the models trained on the different cosmological parameters. As same with test result with different cosmological parameter sets, we used a averaged point from the four subcubes in one simulation box and draw 1,000 points from total 10 runs. 
We see that the 2pcf+FCN result has the largest error, while the CNN result has the smallest error and is also the least biased with respect to truth values.

\begin{figure}[tbp]
\centering 
\includegraphics[width=\textwidth]{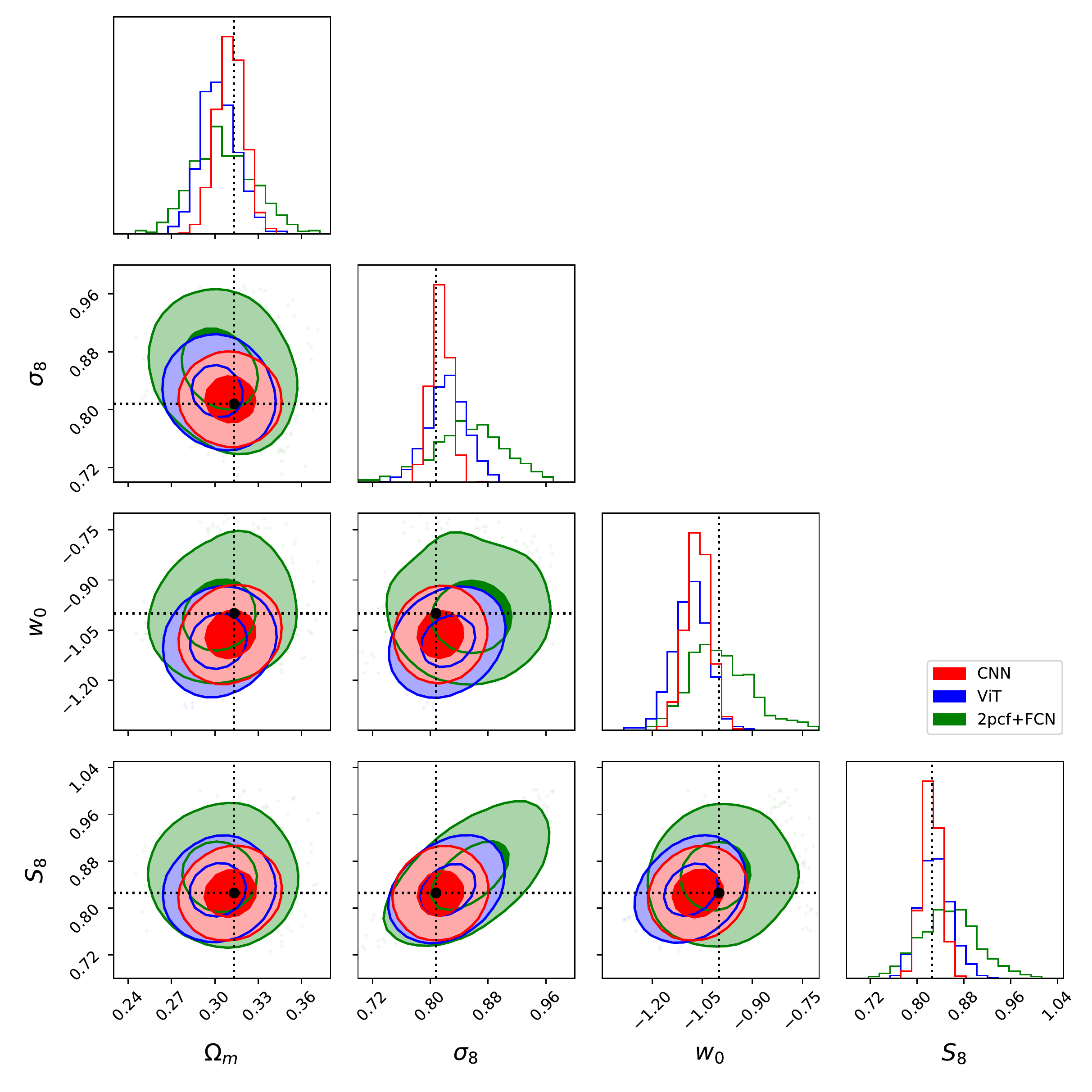}
\caption{\label{fig5} Single cosmology test with CNN (red), ViT (blue), and 2pcf+FCN (green). Black line indicates the truth value of our single cosmology parameter set: $(\Om, \sigma_8, w_0, S_8) = (0.3133, 0.8079, -1, 0.8256)$. The inner line represents 1$\sigma$ and the outer line represents 2$\sigma$.}
\end{figure}

\begin{table}[tbp]
\centering
\begin{tabular}{|lc|c|c|c|}
\hline
\multicolumn{2}{|l|}{}                                         & CNN            & ViT    & 2pcf+FCN            \\ \hline
\multicolumn{1}{|c|}{\multirow{2}{*}{$\Om$}}    & $\left|\rm{mean-truth}\right|$ & \textbf{0.0025} & 0.0103 & 0.0072          \\ \cline{2-5} 
\multicolumn{1}{|c|}{}                       & std             & \textbf{0.0096} & 0.0126 & 0.0213          \\ \hline
\multicolumn{1}{|l|}{\multirow{2}{*}{$\sigma_8$}} & $\left|\rm{mean-truth}\right|$ & \textbf{0.0061} & 0.0161 & 0.0451          \\ \cline{2-5} 
\multicolumn{1}{|l|}{}                       & std             & \textbf{0.0130} & 0.0266 & 0.0489          \\ \hline
\multicolumn{1}{|l|}{\multirow{2}{*}{$w_0$}}  & $\left|\rm{mean-truth}\right|$ & 0.0640          & 0.0845 & \textbf{0.0060} \\ \cline{2-5} 
\multicolumn{1}{|l|}{}                       & std             & \textbf{0.0359} & 0.0516 & 0.0932          \\ \hline
\multicolumn{1}{|l|}{\multirow{2}{*}{$S_8$}}  & $\left|\rm{mean-truth}\right|$ & \textbf{0.0006} & 0.0056 & 0.0301          \\ \cline{2-5} 
\multicolumn{1}{|l|}{}                       & std             & \textbf{0.0153} & 0.0266 & 0.0495          \\ \hline
\end{tabular}
\caption{\label{table2}Absolute value of mean-truth and standard deviation of 100 single cosmology. Bold text is minimum value among 3 methods.
}
\end{table}

When comparing the three methods, we found that the 3D CNN has the lowest standard deviation for each parameter, followed closely by the ViT, while the 2pcf+FCN had consistently the largest statistical scatter, as can be seen in Table~\ref{table2}. We can confirm that CNN, a more prevalent method, can generate accurate results with the lightcone-shaped data containing redshift information. Moreover, we found that ViT can also perform well in this type of scenario. Additionally, the traditional two-point correlation function, combined with a simple dense neural network, performed well.

\subsection{Correlation and Combined Results}
\label{sec4.2}

\begin{figure}[tbp]
\centering 
\includegraphics[width=\textwidth]{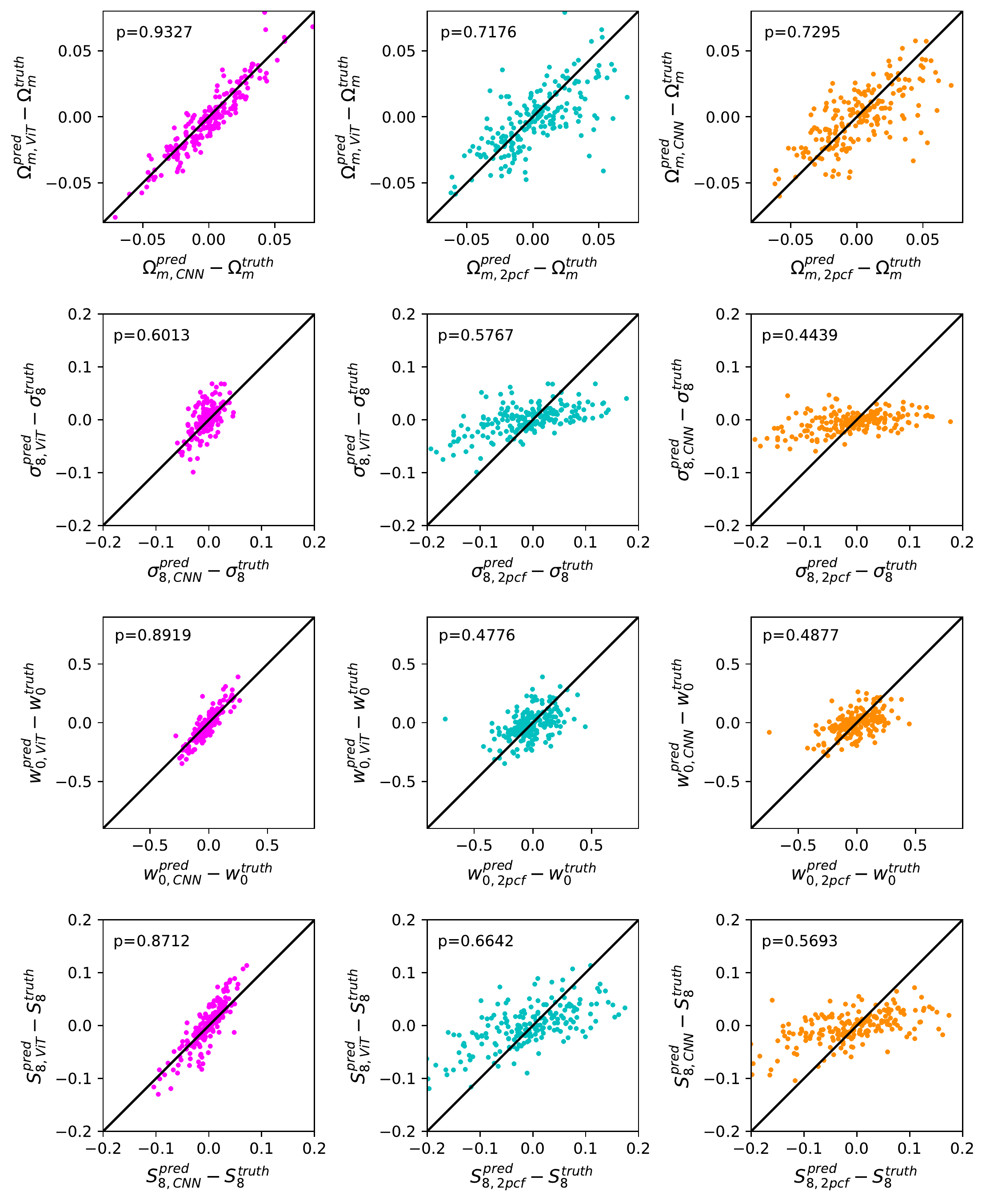}
\caption{\label{fig6} Relations between the difference between predictions and truth values from our three models. From top to bottom panels: $\Om$, $\sigma_8$, $w_0$, and $S_8$. From left to right panels: CNN vs. ViT, 2pcf+FCN vs. ViT, and 2pcf+FCN vs. CNN. 
The Pearson coefficient, as computed from Eq.~\ref{eq4}, for the data is displayed in the top left of each panel.}
\end{figure}

To assess the relationship between the three methods, we employ Pearson correlation to the differences between the predictions and truth values of each of the four parameters from the different methods. 
We calculate the Pearson correlation using 
\begin{equation}\label{eq4}
p = \frac{{\rm Mean}[(X-\mu_X)(Y-\mu_Y)]}{\sigma_X \sigma_Y}.
\end{equation}
where $X$ and $Y$ are groups of parameters we want to compare, and $\mu_{X,Y}$ and $\sigma_{X,Y}$ are their mean and standard deviation, respectively. 
When $p$ is close to $+1$ or $-1$, it means that $X$ and $Y$ are linearly correlated or anti-correlated, respectively. On the other hand, when the Pearson correlation is close to 0, it means there is no linear correlation between $X$ and $Y$. 
From Figure~\ref{fig6}, we can observe that the results from CNN and ViT have relatively higher linear correlations than 2pcf+FCN has with others. This suggests that averaging the predictions from CNN/ViT with 2pcf+FCN could yield better results.
Although the predictions for $\sigma_8$ between CNN and ViT are not strongly correlated and thus the two models may be using different parts of the density field to arrive at their results. Taken together, these results suggest that averaging all 3 models may provide the best cosmological constraints.

\begin{figure}[tbp]
\centering 
\includegraphics[width=\textwidth]{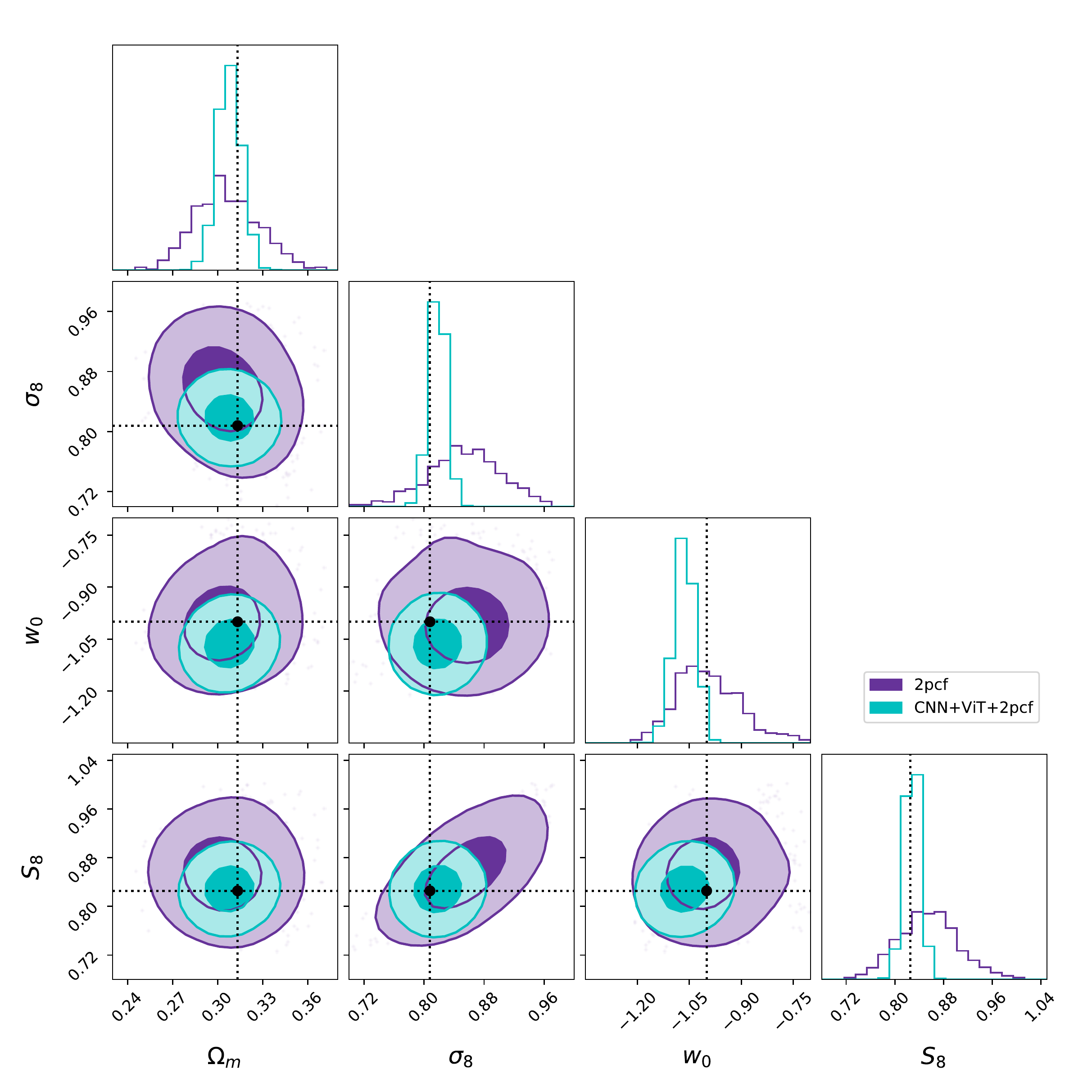}
\caption{\label{fig7} Same as Figure~\ref{fig5}, except 2pcf+FCN (purple) 
and combined results from CNN, ViT, and 2pcf+FCN (cyan) are drawn.}
\end{figure}

\begin{table}[tbp]
\centering
\begin{tabular}{|lc|c|c|c|}
\hline
\multicolumn{2}{|l|}{}                                         & 2pcf+FCN            & (CNN, ViT)    & (CNN, ViT, 2pcf+FCN)           \\ \hline
\multicolumn{1}{|c|}{\multirow{2}{*}{$\Om$}}    & $\left|\rm{mean-truth}\right|$ & 0.0072 & \textbf{0.0054} & 0.0056          \\ \cline{2-5} 
\multicolumn{1}{|c|}{}                       & std             & 0.0213 & 0.0082 & \textbf{0.0080}          \\ \hline
\multicolumn{1}{|l|}{\multirow{2}{*}{$\sigma_8$}} & $\left|\rm{mean-truth}\right|$ & 0.0451 & \textbf{0.0081} & 0.0101          \\ \cline{2-5} 
\multicolumn{1}{|l|}{}                       & std             & 0.0489 & 0.0116 & \textbf{0.0112}          \\ \hline
\multicolumn{1}{|l|}{\multirow{2}{*}{$w_0$}}  & $\left|\rm{mean-truth}\right|$ & \textbf{0.0060}          & 0.0707 & 0.0637 \\ \cline{2-5} 
\multicolumn{1}{|l|}{}                       & std             & 0.0932 & 0.0316 & \textbf{0.0312}          \\ \hline
\multicolumn{1}{|l|}{\multirow{2}{*}{$S_8$}}  & $\left|\rm{mean-truth}\right|$ & 0.0301 & \textbf{0.0009} & 0.0029          \\ \cline{2-5} 
\multicolumn{1}{|l|}{}                       & std             & 0.0495 & 0.0134 & \textbf{0.0129}          \\ \hline
\end{tabular}
\caption{\label{table3} Same to Table~\ref{table2}, except showing 2pcf+FCN, the combined results from CNN and ViT, and the combined results from all three models. See texts for how to calculate the weighted mean.}

\end{table}

Next, we study the performance of combined results of two or more models by calculating  the weighted mean 

\begin{equation}\label{eq5}
\bar{x}_{i, \mathcal{M}} = \frac{\displaystyle \sum_{M \in \mathcal{M}} w_{i, M} x_{i, M} }{\displaystyle \sum_{M \in \mathcal{M}} w_{i, M} } ~,
\end{equation}
where $x_{i,M}$ is a prediction of the given cosmological parameter from  a model $M$, which is a member of our combined model set $\mathcal{M}$, and $w_{i,M} \equiv \sigma_{i,M}^{-2}$ is its weight defined as the inverse of the parameter variance calculated from 10 runs of the single cosmology test. 
The combined result of CNN, ViT, and 2pcf+FCN is displayed in Figure~\ref{fig7} compared with the 2pcf+FCN alone. In Table~\ref{table3}, we show the absolute deviation and $1$-$\sigma$ error for each parameter for the 2pcf+FCN alone and for the combined models of (CNN, ViT) and (CNN, ViT, 2pcf+FCN). As expected, the combination of all three models, (CNN, ViT, 2pcf+FCN), provide the best parameter estimation for most cases, where the standard deviation is reduced $1-4\%$ from (CNN, ViT)-combination and $60-77\%$ from 2pcf+FCN alone. Also, mostly thanks to CNN, (CNN, ViT, 2pcf+FCN)-model predict $\sigma_8$ and $S_8$ parameter without noticeable degeneracy (see $\sigma_8$-$S_8$ panel in Figure~\ref{fig7}).

\subsection{Importance Maps}
\label{sec4.3}
Understanding the inner workings of machine learning algorithms is crucial when applying deep learning techniques in scientific research. One popular method for this introspection is Gradient-weighted Class Activation Mapping (Grad-CAM) \cite{gradcam}. Grad-CAM computes weights by utilizing gradients through backpropagation between the last layer of the FCN and the last convolutional layer.
 
We can obtain weights $\alpha^c_k$ for each kernel, $k$, and class $c$ by calculating the gradients of $y_c$, the classification score before softmax activation, with respect to $A^{k}_{ij}$, the feature maps of the last convolution layer where $i, j$ are spatial indices of the 2D image. These weights are thus given by,
\begin{equation}\label{alpah}
\alpha^c_k = \frac{1}{Z}\sum\limits_{i}\sum\limits_{j}\frac{\partial y_c}{\partial A^{k}_{ij}},
\end{equation}
where $Z$ is a weighted sum of feature maps and the global average pooling is performed by adding up the gradients for each $i, j$ component and then dividing the sum by $Z$. 
After obtaining $\alpha^c_k$, they are multiplied by the feature maps of their corresponding convolutional layer. In classification problems, higher values of $y_c$ indicate more accurate classification, so the ReLU function is applied to show only positive gradients as $y_c$ increases. This calculation yields
\begin{equation}
    L^c_{Grad-CAM} = ReLU(\sum\limits_{k} \alpha^c_k A^k),
\end{equation}
which is final Grad-CAM outputs.

However, in our problem, which is a regression task, some modifications are necessary before applying Grad-CAM. Our final layer represents the actual values instead of class probabilities. Therefore, solely considering high gradients is not appropriate for our objective. To address this, we referred to \cite{gradcamforregression} where, in the case of a regression task, $d=1/D$ was introduced to replace $y_c$, where $D=\sqrt{(x-x')^2}$ and indicates the amount of deviation between the prediction, $x$, and the truth value $x'$. Smaller $D$ values indicate higher accuracy, and by taking the inverse, we can achieve the desired behavior where $d$ increases as the accuracy improves.
The equation ~\ref{alpah}, the gradient of $y^c$, changes to become the gradient of $d$,
\begin{equation}
    \alpha^d_k = \frac{1}{Z}\sum\limits_{i}\sum\limits_{j}\frac{\partial d}{\partial A^{k}_{ij}},
\end{equation}
where $\frac{\partial d}{\partial A^{k}_{ij}} = \frac{\partial d}{\partial x} \frac{\partial x}{\partial A^{k}_{ij}}$, and $\frac{\partial d}{\partial x} = \frac{-1}{(x-x')^2}$ via the chain rule. Taking  $(x-x')^2$ to be a constant, shows that we need only to consider negative gradients of the prediction to adequately amend the original Grad-CAM equation.

While Grad-CAM was originally designed for CNNs, its underlying principle of using gradients to understand feature importance can be adapted for various neural network architectures, including attention-based models like Vision Transformers. For CNN, we use last convolution layer after activation and for ViT, we utilize the last normalized layer after the transformer calculation is done to obtain the gradients. We use the weights trained by our training dataset.

\begin{figure}[h]
\centering
\includegraphics[width=\linewidth]{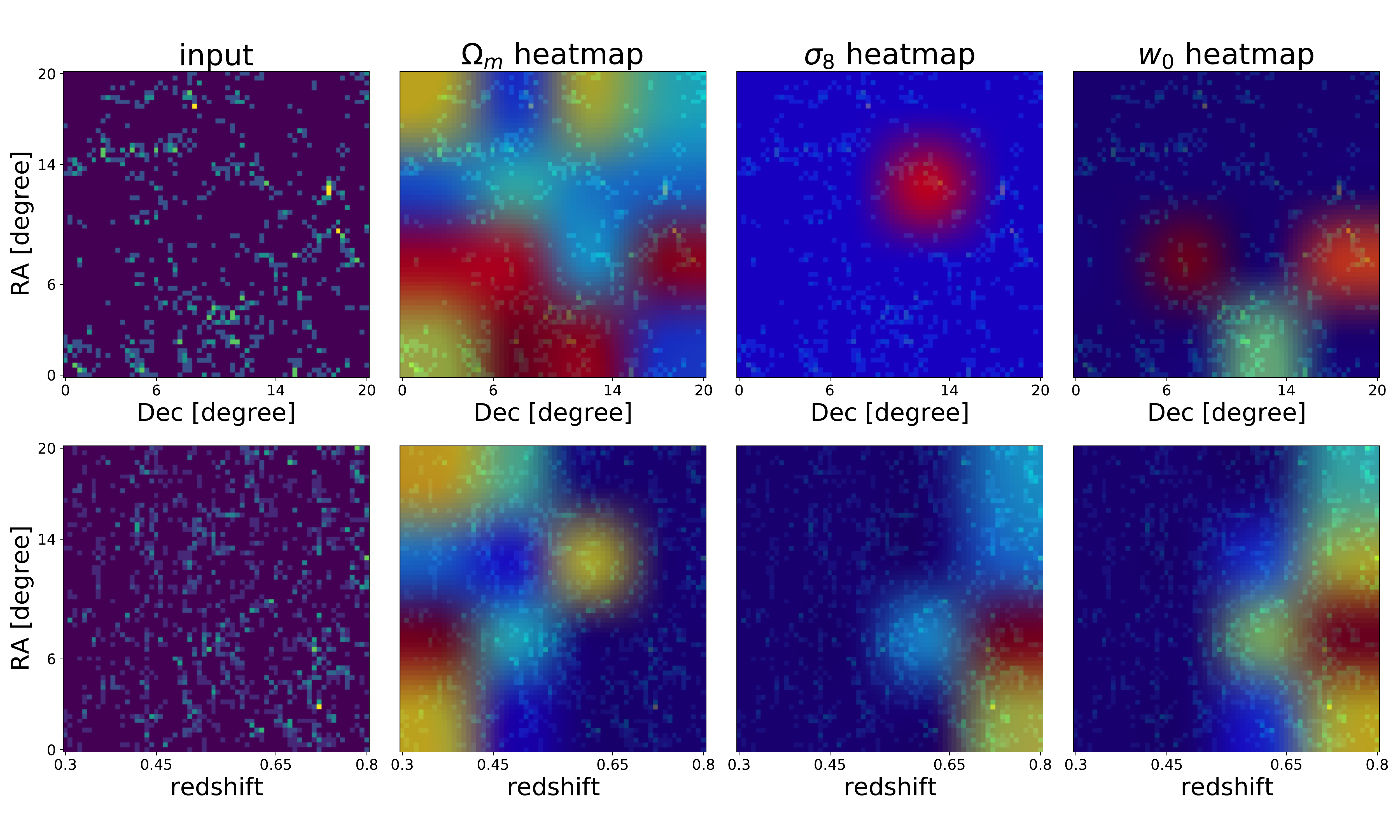}
\caption{\label{fig8} Grad-CAM map of Convolution Neural Network. (Top) Grad-CAM map from RA-Dec plane. (Bottom) Grad-CAM map from RA-redshift plane.
}
\end{figure}

\begin{figure}[h]
\centering
\includegraphics[width=\linewidth]{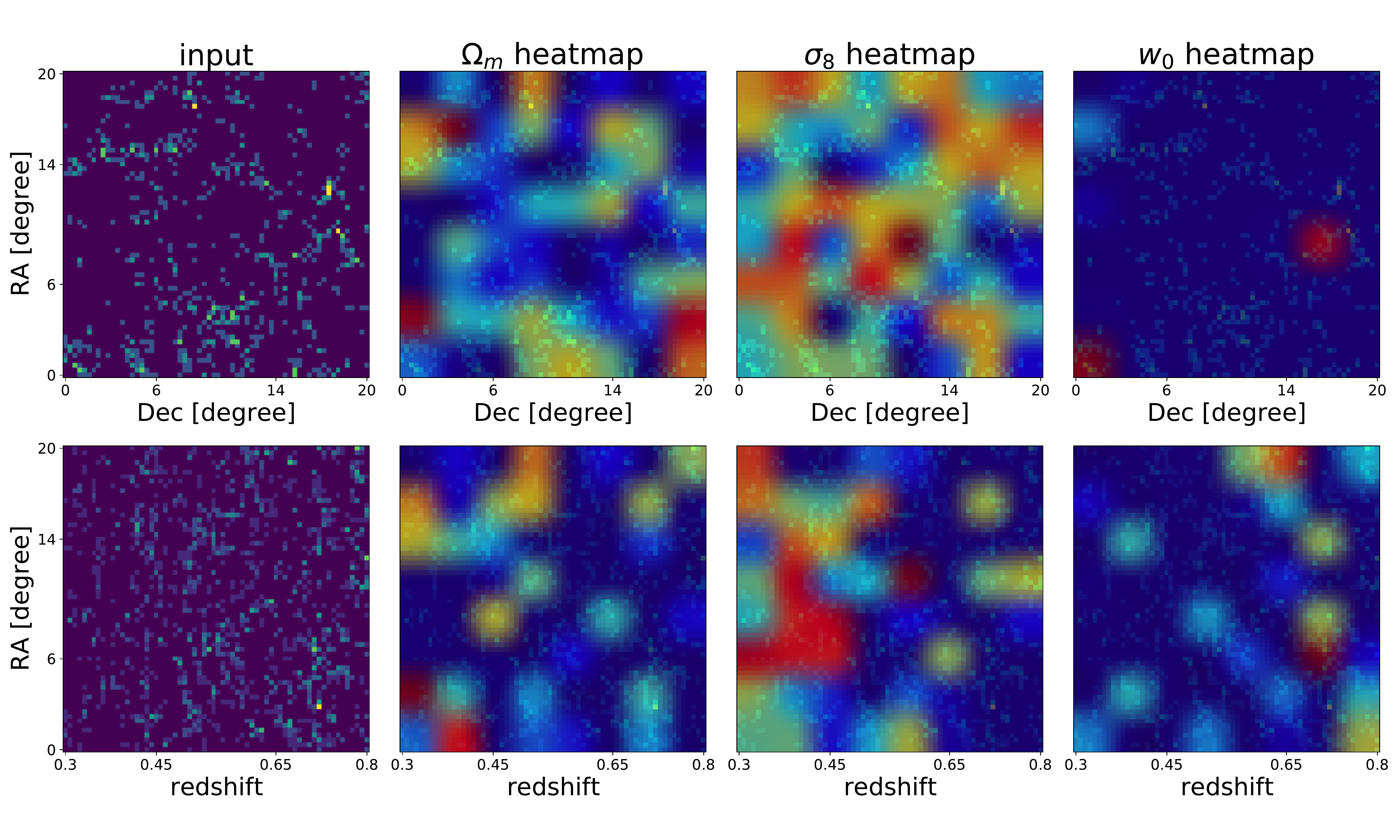}
\caption{\label{fig9} Grad-CAM map of Vision Transformer. (Top) Grad-CAM map from RA-Dec plane. (Bottom) Grad-CAM map from RA-redshift plane.
}
\end{figure}
Figures ~\ref{fig8} and Figure~\ref{fig9} show a Grad-CAM heatmap of a sample single cosmology data for a CNN model and a ViT model, respectively. The red color which means high Grad-CAM score indicates that the corresponding region contains essential information, while blue color signifies less important parts. 

The final Grad-CAM result for CNN has a shape of $4^3$, which corresponds to the shape before the last average pooling, while ViT maintains the original patch size of $8^3$ throughout the algorithm's calculations. This is why the colored box size in two plots look different.

\begin{figure}[tbp]
\centering
\includegraphics[width=\linewidth]{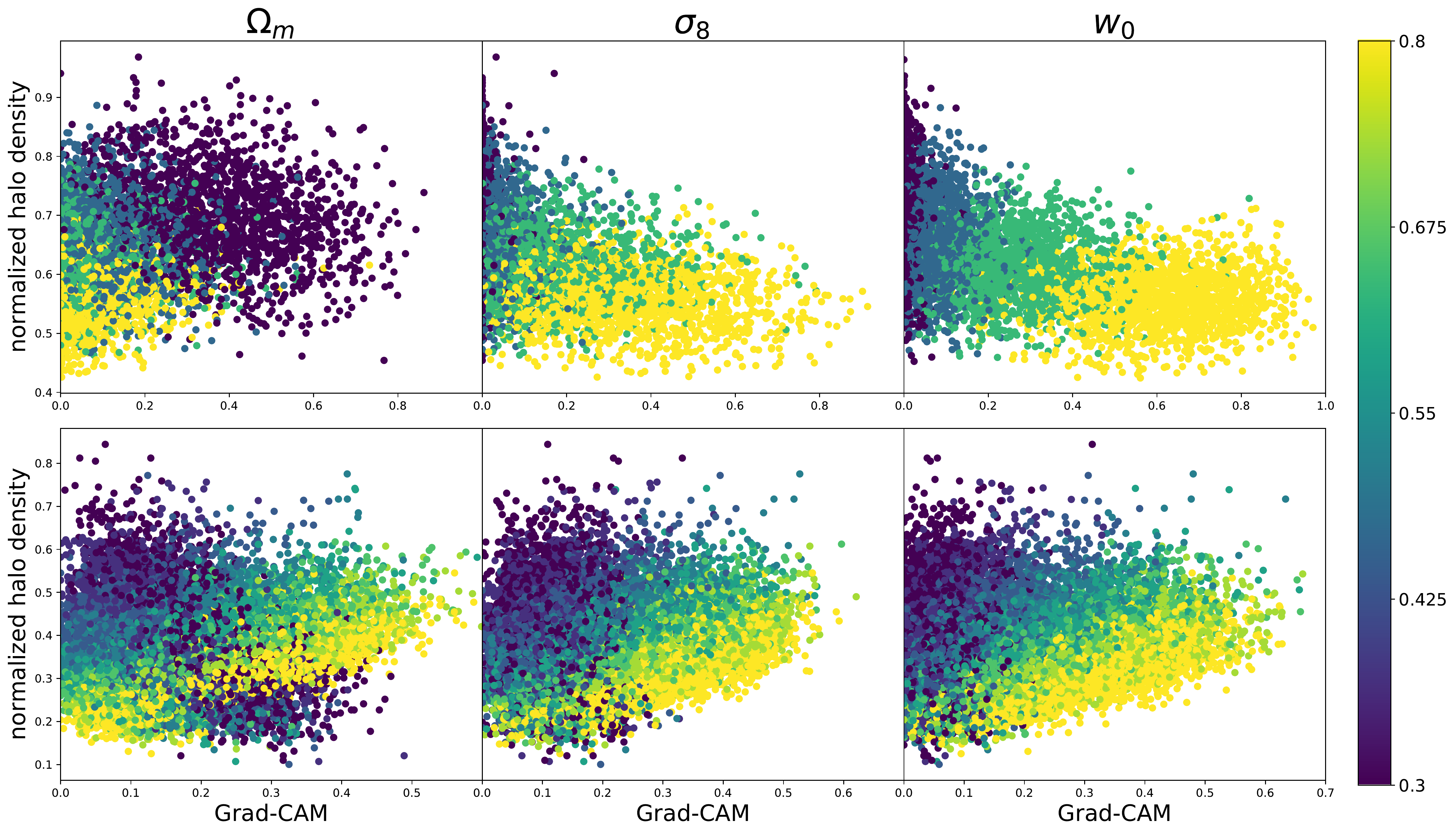}
\caption{\label{fig10} The correlation between normalized halo density and Grad-CAM value ofsingle cosmology catalogues, while using CNN (top) and ViT (bottom). The color bar indicates the redshift of the pixel. 
}
\end{figure}

A visual comparison between the Grad-CAM importance maps and the input data is difficult to interpret by eye. In figure~\ref{fig10} we match pixels importance to their input density value, while also colour coding for the pixel's input redshift. We generated this figure from the single cosmology data and used 100 subcubes for CNN and 40 subcubes for ViT since the final Grad-CAM output shapes of the two algorithms differ, which results in a different number of data points.

For CNN (top panels) we see that the model prediction of $\Omega_m$ is influenced most predominately by over densities at lower redshift, while $\sigma_8$ and $w_0$ are more related to under-dense and higher redshift pixels. In the case of ViT (lower panels), there is an upward trend diagonally across all three parameters. This indicates that as the halo density increases at high redshifts, Grad-CAM values also tend to increase. Specifically, for $\Omega_m$, it is noticeable that the halo density at low redshifts has a more significant impact compared to other parameters.

We adopted this approach to gain a better understanding of where our algorithm was focusing within our input data. For example, in the Grad-CAM paper~\cite{gradcam}, when classifying a ``tiger cat'', Grad-CAM indicated a high concentration on the stripes of the cat's body, while for classifying a ``dog'', it focused on the dog's face. The reason we can interpret these findings in this manner is that as humans, we can easily differentiate between dogs and cats when viewing the images, and we can directly judge whether Grad-CAM's results are reasonable or not. However, in our case with a halo density field, it is challenging to immediately determine which areas to focus on for parameter prediction. Therefore, we examined the relationship between the normalized halo density and Grad-CAM.

\section{Conclusions}
\label{sec:conc}

Predicting cosmological parameters from large-scale structures is a crucial aspect of cosmology and a field that benefits greatly from machine learning and big data techniques. In this study, we estimated three cosmological parameters, $\Om$, $\sigma_8$, $w_0$, and one derived parameter, $S_8$, using two deep learning algorithms --- Convolutional Neural Network (CNN) and Vision Transformer (ViT) --- for the first time. We also compared these results with a statistical approach that combined standard two-point correlation functions and a simple neural network regression. Our comparison revealed that CNN currently yield the best results, while ViT also show significant potential when applied to predicting cosmological parameters.

Using the same data, we found that the combination of ViT, CNN, and 2pcf resulted in a 62\% reduction in the  error of $\Om$, a 
77\% reduction in the error of $\sigma_8$, a 66\% reduction in the error of $w_0$, and a 74\% reduction in the error of $S_8$, compared with the 2pcf+FCN alone.

We have shown that ViT could play a role in cosmological model constraints, but they may benefit from pre-training on other datasets or using transfer learning for better performance. Leveraging deep learning method makes us constrain cosmological parameters more tightly than using 2pcf with simple fully connected layers. This proof-of-concept work could be made applicable to current data by forward modeling observational systematics to mock particular cosmological surveys, e.g., SDSS, DESI, etc. We plan to undertake this approach in our future work.

To gain deeper insights into how our machine learning algorithms interpret large scale structure data, we utilized the Gradient-weighted Class Activation Mapping (Grad-CAM) technique. This method enabled us to identify which regions of the data were most informative for the models. Our analysis revealed that Convolutional Neural Networks (CNNs) and Vision Transformers (ViTs) extract valuable information from distinct areas within the dataset. Furthermore, we found that even within a single algorithm, the focus shifts depending on which cosmological parameter is being predicted. Upon statistical evaluation, we also noted variations in the Grad-CAM values across different redshifts and halo densities.

Distinguished among studies that employ machine learning for the prediction of cosmological parameters, our work represents an innovative effort that delves into the inner workings of these algorithms. Rather than presenting our results as definitive answers, our primary objectives were to compare information content between various methods and to gain a human perspective on how our these algorithms accentuated particular facets of the data to yield results. Our approach acknowledges that the outcomes are contingent upon the choice of data and the specific machine learning algorithms and architectures employed.

\acknowledgments

We would like to thank Dongsu Bak, Sangnam Park, Ian Watson and Hannah Jhee for helpful discussions and considerate comments.
Also we thanks to ChatGPT for valuable feedback and editorial input.

This research was supported by Basic Science Research Program through the National Research Foundation of Korea (NRF) funded by the Ministry of Education (2018\-R1\-A6\-A1\-A06024977).
C.G.S is also support via the Basic Science Research Program from the National Research Foundation of South Korea (NRF) funded by the Ministry of Education (2020\-R1\-I1\-A1\-A01073494).
S.E.H. was supported by the project \begin{CJK}{UTF8}{mj}우주거대구조를 이용한 암흑우주 연구\end{CJK} (``Understanding Dark Universe Using Large Scale Structure of the Universe''), funded by the Ministry of Science. 
Our research was partially carried out using the University of Seoul UBAI big data cluster. 
The authors are especially grateful to KISTI for providing the KREONET network.

We would like to acknowledge the Korean Astronomy and Machine Learning (KAML) meeting series for providing a forum for fruitful discussions on the intersection of astronomy and machine learning.

\bibliographystyle{JHEP}
\bibliography{ref}

\end{document}